\documentclass[aps,showpacs,preprintnumbers,showkeys,eqsecnum,amsmath,amssymb]{revtex4}

\usepackage{float}
\usepackage{graphics,epsfig}
\usepackage{graphicx}
\usepackage{dcolumn}
\usepackage{bm}

\begin{document}
\baselineskip=0.8 cm

\title{ Analytical expressions for greybody factor and dynamic evolution \\
for scalar field in Ho\v{r}ava-Lifshitz black hole}
\author{Chikun Ding }
\author{Songbai Chen}
\author{Jiliang {Jing}\footnote{Corresponding author, Electronic address:
jljing@hunnu.edu.cn}} \affiliation{ Institute of Physics and
Department of Physics, Hunan Normal University, Changsha, Hunan
410081, People's Republic of China \\ and \\ Key Laboratory of Low
Dimensional Quantum Structures and Quantum Control (Hunan Normal
University), Ministry of Education, P. R. China.}

\vspace*{0.2cm}
\begin{abstract}
\baselineskip=0.6 cm

\begin{center}
{\bf Abstract}

\end{center}

We investigate the propagation and evolution for a massless scalar
field in the background of $\lambda=1/2$ Ho\v{r}ava-Lifshitz black
hole with the condition of detailed balance. We fortunately obtain
an exact solution for the Klein-Gordon equation. Then, we find an
analytical expression for the greybody factor which is valid for any
frequency; and also exactly show  that the perturbation decays
without any oscillation. All of these can help us to understand more
about the Ho\v{r}ava-Lifshitz gravity.

\vspace*{0.2cm}
\end{abstract}


\maketitle
\newpage

\section{introduction}

Einstein's general relativity is increasingly important in the
modern physics,  especially in the frontiers of very large distance
scales including astrophysics and cosmology. However, it was proved
to be nonrenormalizable by quantum field theories. In other words,
the study of the ultra-violet (\textrm{UV}) completion of gravity
has been a difficult road for theoretical physics in the past 50
years. The only convincing answer may be string theory, but it works
only in perturbation theory and at energies well below the
 Plank scale. Recently, Ho\v{r}ava \cite{ho1} proposed a different field
 theory model for a \textrm{UV} complete theory of gravity which can be a power
counting renormalizable gravity theory in four dimensions. The
Ho\v{r}ava-Lifshitz theory is based on the perspective that Lorentz
symmetry should appear as an emergent symmetry at long distances,
but can be fundamentally absent at high energies \cite{Pav}. So
Ho\v{r}ava considered a system whose scaling at short distance
exhibits a strong anisotropy between space and time. In the infrared
limit, the higher derivative terms do not contribute and it reduces
to standard general relativity. Thus, the Ho\v{r}ava-Lifshitz
gravity can be regarded as an ultraviolet complete theory of
general relativity.

Because of these novel features, the Ho\v{r}ava-Lifshitz gravity has
been intensively investigated \cite{ho2,ho3,VW,klu,Nik,Vol,CH,CHZ}
and its cosmological applications have been  studied
\cite{cal,TS,muk,Bra,KK} since then. Some metrics of the static
spherically symmetric black holes with nonvanishing cosmological
constant have been obtained in the Ho\v{r}ava-Lifshitz theory
\cite{CY,LMP,CCO,CLS,Gho} and the associated thermodynamic
properties of those black holes have been investigated
\cite{MK,Nis,CCO1,Myung}. It is well known that the
Ho\v{r}ava-Lifshitz gravity has several free parameters which should
be fixed to obtain general relativity at large scales. According to
Ref. \cite{LMP}, the general covariance is restored only for
coupling constant $\lambda=1$, and for $\lambda \neq 1$ the
Ho\v{r}ava-Lifshitz gravity seems not to reduce to general
relativity even in large scales. Thus, the potentially observable
properties of black holes in the deformed Ho\v{r}ava-Lifshitz
gravity have been considered by using the gravitational lensing
\cite{sb1,RAK1}, quasinormal modes \cite{sb2} and the accretion disk
\cite{disk}, {\it et al}..

To understand  the properties of the Ho\v{r}ava-Lifshitz theory, it
is necessary to  study other cases in which $\lambda\neq 1$. The
main purpose of this paper is to study the propagation and dynamical
evolution of a massless scalar field in the Ho\v{r}ava-Lifshitz
black-hole with the coupling constant $\lambda=\frac{1}{2}$.  The
reason we take  $\lambda=\frac{1}{2}$ here is that an exact solution
for the Klein-Gordon equation in the spacetime can be obtained. Then
the greybody factor and quasinormal modes can be worked out
analytically.

The paper is organized as follows. In Sec. II we present the
black-hole solutions in the Ho\v{r}ava-Lifshitz gravity. In Sec.
III, by exact analytical calculation, we study the greybody of the
massless scalar field propagating in the Ho\v{r}ava-Lifshitz
black-hole spacetime with the coupling constant
$\lambda=\frac{1}{2}$. In Sec. IV, we study its dynamical evolution
of the scalar field. In the last section, we summarize and discuss
our conclusions

\section{The black holes in the Ho\v{r}ava-Lifshitz gravity}

The four-dimensional metric in the ADM formalism can be expressed as
\cite{adm}
\begin{eqnarray}
 ds_{ADM}^2= - N^2 dt^2 + g_{ij} \Big(dx^i - N^i dt\Big)
\Big(dx^j - N^j dt\Big),
\end{eqnarray}
and the action of the nonrelativistic renormalizable gravitational
theory proposed by Ho\v{r}ava is given by
\begin{eqnarray}\label{action2}
S_{HL}&=&\int dtd^3x, \Big({\cal L}_0 + {\cal L}_1\Big) ,
\end{eqnarray}
with
\begin{eqnarray}
{\cal L}_0 &=& \sqrt{g}N\left\{\frac{2}{\kappa^2}(K_{ij}K^{ij}
\label{action1}-\lambda K^2)+\frac{\kappa^2\mu^2(\Lambda_W R
  -3\Lambda_W^2)}{8(1-3\lambda)}\right\}\,,\nonumber \\ {\cal L}_1&=&
\sqrt{g}N\left\{\frac{\kappa^2\mu^2 (1-4\lambda)}{32(1-3\lambda)}R^2
-\frac{\kappa^2}{2w^4} \left(C_{ij} -\frac{\mu w^2}{2}R_{ij}\right)
\left(C^{ij} -\frac{\mu w^2}{2}R^{ij}\right) \right\}, \nonumber
\end{eqnarray}
where $\kappa^2$, $\mu$, $\Lambda$, and $\omega$  are constant
parameters, $N^i$ is the shift vector, $K_{ij}$  is the extrinsic
curvature, and $C_{ij}$ the Cotten tensor defined by $
C^{ij}=\epsilon^{ik\ell}\nabla_k\big(R^j{}_\ell
-\frac14R\delta_\ell^j\big)=\epsilon^{ik\ell}\nabla_k R^j{}_\ell
-\frac14\epsilon^{ikj}\partial_kR\,.\label{def.K.C} $ Taking
$N^i=0$, the spherically symmetric solution is given by~\cite{LMP}
\begin{eqnarray}
\label{ssm} ds_{SS}^2 = - \tilde{N}^2(r)f(r)\,dt^2 +
\frac{dr^2}{f(r)} + r^2 (d\theta^2 +\sin^2\theta d\phi^2), \,
\end{eqnarray}
with
\begin{eqnarray}
\label{sol1}\label{sol2}
\tilde{N}&=&(\mathrm{R}r)^{q_\pm(\lambda)},~~~~q_\pm(\lambda)=
-\frac{1+3\lambda\pm2\sqrt{6\lambda-2}}{\lambda-1} \,, \nonumber  \\
f&=&1 + (\mathrm{R}r)^2-m
(\mathrm{R}r)^{\frac{1-q}{2}},
\end{eqnarray}
where $\mathrm{R}=\sqrt{-\Lambda_W}$, and $m$ is an integration
constant related to the mass. Hereafter we choose the negative sign
of $q_{\pm}(\lambda)$ because the metric for $q_{+}(\lambda)$ is no
physical meaning. Thus, the line element (\ref{ssm}) becomes
\begin{eqnarray}\label{metric05}
ds^2=-(\mathrm{R}r)^{2\xi} f(r)dt^2+\frac{1}{f(r)}dr^2+r^2d\Omega^2,
\end{eqnarray}
with\begin{eqnarray}\label{add1}
 f=1-m(\mathrm{R}r)^{\frac{1-\xi}{2}}+(\mathrm{R}r)^2, \;\;
\end{eqnarray}where $\xi=q_-(\lambda)$, and its Hawking temperature is \begin{eqnarray}
 T_H=\frac{\mathrm{R}}{4\pi}\Big[2(\mathrm{R}r_H)^{\xi+1}-\frac{1-\xi}{2}m
 (\mathrm{R}r_H)^{(\xi-1)/2}\Big].
\end{eqnarray}

Though the broken of the Lorentz symmetry arises in small scale in
the Ho\v{r}ava-Lifshitz theory, the notions of a black hole in the
Ho\v{r}ava-Lifshitz theory in the large scale may also be defined as
that in the Einstein's general relativity \cite{ding}. Here we
calculate its curvature scalar involving cosmological constant
\begin{eqnarray}
R_{\mu\nu\tau\rho}R^{\mu\nu\tau\rho}&=&4\mathrm{R}^4\big[\xi^4+4\xi^3+8\xi^2+8\xi+6\big]
+\frac{4\xi} {r^4}\big[\xi(\xi^2-2\xi+3)+2\mathrm{R}^2r^2(\xi+1)(\xi^2+1)\big]
\nonumber\\
&&-\frac{m}{r^4}(\mathrm{R}r_H)^{(1-\xi)/2}\big[\xi(3\xi+1)(\xi^2-2\xi+5)
+\mathrm{R}^2r^2(3\xi^4+4\xi^3+10\xi^2+8\xi+15)\big]
\nonumber\\
&&+\frac{3m^2\mathrm{R}}{16r^3}(\mathrm{R}r)^{-\xi}(3\xi^2+2\xi+3)(\xi^2-2\xi+9).
\end{eqnarray}
Which shows that $r=0$ is an intrinsic singularity and $r=r_H$ is a
coordinate singularity in the Ho\v{r}ava-Lifshitz spacetime
(\ref{metric05}). Therefore, we can still define a black hole in
which the event horizon is identified as the coordinate singularity.

\section{Greybody factor of the scalar field propagating
in the Ho\v{r}ava-Lifshitz black-hole spacetime}

Consider a massless scalar field $\Phi$ propagating in the
background of the Ho\v{r}ava-Lifshitz black hole and let this mode has definite frequency
$\omega$, so we can set
 \begin{eqnarray}\Phi(t,r,\theta,\varphi)=e^{i\omega
t}\phi(r)Y_{lm}(\theta,\varphi).\end{eqnarray}
The equation of a
massless scalar field is
\begin{eqnarray}\label{k-g}
\frac{1}{\sqrt{-g}}\partial_\mu(\sqrt{-g}
g^{\mu\nu}\partial_\nu)\Phi(t,r,\theta,\varphi)=0.
\end{eqnarray}  Defining a tortoise
coordinate
\begin{eqnarray}\label{tortoise}dx=\frac{dr}{F(r)},\end{eqnarray}
where $F(r)=(\mathrm{R}r)^\xi f(r)$, we find that the radial equation for
the scalar field reads
\begin{eqnarray}\label{schrodinger}
\bigg(\frac{d^2}{dx^2}+\omega^2-V\bigg)(r\phi)=0,
\end{eqnarray}
with
\begin{eqnarray} V=F(r)\bigg[\frac{1}{r}\frac{d F(r)}{dr}
+\frac{l(l+1)}{r}{\mathrm{R}}\bigg],\nonumber
\end{eqnarray}
where $l$ is the angular quantum number.

\subsection{The boundary conditions}

In the near-horizon region ($r\simeq r_H$ and $V(r)\ll \omega^2$)
the purely ingoing solution can be expressed
\begin{eqnarray}
\phi(r) = A e^{i \omega x}. \label{solhor}
\end{eqnarray}
Since Eq. (\ref{schrodinger}) takes  Schr\"odinger-like form, the
conversation current can be defined as
\begin{eqnarray}
j = \frac{1}{2i} \left( \phi^* \frac{d\phi}{dx} - \phi
\frac{d\phi^*}{dx} \right). \label{flux}
\end{eqnarray}
The current is nothing but the flux per unit  coordinate area. Thus,
the flux  near the horizon is
\begin{eqnarray}
J_{hor} = 4\pi r_H^2 j|_{r_H}=4\pi r_H^2\omega |A|^2   \ .
\label{flux-hor}
\end{eqnarray}

Now we consider the asymptotic infinity boundary condition. For
large $r$ (i.e. $r \gg r_H$), the metric function in
(\ref{metric05}) becomes $f(r)\sim(\mathrm{R}r)^2$
 since $\frac{1-\xi}{2}<2$ for any $\lambda$, then Eq. (\ref{k-g})
reduces to
\begin{eqnarray} \label{boundary1}
\left[r^2 \frac{d^2}{dr^2} + (\xi+4)r \frac{d}{dr} +
\frac{\omega^2}{\mathrm{R}^2(\mathrm{R}r)^{2(\xi+1)}}-\frac{l(l+1)}{\mathrm{R}^2r^2} \right]
\phi(r) = 0  .
\end{eqnarray}
We find the equation (\ref{k-g}) has a pair of exact solutions when
$\xi=1$ (with coupling constant $\lambda=1/2$). So we focus our
attention on the case of $\xi=1$. Let
 \begin{eqnarray}
&&u=\frac{\omega}
{2\mathrm{R}^3r^2},\ \nonumber \\
&&\phi(r(u))=u\mathcal{F}(u),\nonumber
\\
&&\tilde{L}^2=\frac{l(l+1)}{\mathrm{R}^2r_H^2}, \nonumber  \\&&
\tilde{\omega}=\frac{\omega}{\mathrm{R}^3r_H^2}=\frac{\omega}{2\pi
T_H},
\end{eqnarray}
Eq. (\ref{boundary1}) becomes
 \begin{eqnarray}\label{boundaryeq}
u^2\frac{d^2\mathcal{F}}{du^2}+u\frac{d\mathcal{F}}{du}
+\Big[u^2-1-\frac{\tilde{L}^2}{2\tilde{\omega}}u\Big]\mathcal{F}=0.
\end{eqnarray}
Eq. (\ref{boundaryeq}) can be transformed into a confluent
hypergeometric equation, and its solutions can be represented by the
first and second Kummer's functions. Then the boundary wave function
$\phi$ is
\begin{eqnarray}\label{boundarysol}
\phi(r(u))&=&e^{-iu}u^2\Big[\hat{C}_1M(\hat{a},3,2iu)
+\hat{C}_2U(\hat{a},3,2iu)\Big], \end{eqnarray} with
\begin{eqnarray}
\hat{a}&=&\frac{3}{2}-\frac{i\tilde{L}^2}{4\tilde{\omega}},
\end{eqnarray}
where $\hat{C}_1$ and $\hat{C}_2$ are constants. Using the
analytical expression of the Kummer's functions under the condition
of $u\rightarrow0$
\begin{eqnarray}
&&M(\hat{a},3,2iu)= \sum_{n=0}^\infty\frac{(\hat{a})_n}{(3)_nn!}(2iu)^n, \nonumber \\
&&U(\hat{a},3,2iu)=-\frac{1}{4\Gamma(\hat{a})}u^{-2}
+\frac{i(\hat{a}-2)}{2\Gamma(\hat{a})}u^{-1}
-\frac{1}{2\Gamma(\hat{a}-2)}\nonumber \\
&&\qquad\qquad\qquad\cdot
\sum_{n=0}^\infty\frac{(\hat{a})_n(2iu)^n}{(3)_nn!}\Big[\ln(2iu)+
\psi(\hat{a}+n)-\psi(1+n)-\psi(3+n)\Big],
\end{eqnarray}
with $(\rho)_0=1,\  (\rho)_n=\Gamma(\rho+n)/\Gamma(\rho),\
\psi(\rho)=d\ln\Gamma(\rho)/d\rho$,
the far-region boundary wave (\ref{boundarysol}) becomes
\begin{eqnarray}\label{sol_asy2} &&\phi(u)= \hat{C}_1e^{-iu}u^2\sum_{n=0}^\infty\frac{(\hat{a})_n}{(3)_n}(2iu)^n
-\frac{\hat{C}_2}{4\Gamma(\hat{a})} \nonumber \\
&&\qquad\qquad\cdot\left\{
1-\frac{\tilde{L}^2}{2\tilde{\omega}}u
+\Big(\frac{1}{2}+\frac{i\tilde{L}^2}{2\tilde{\omega}}\Big)u^2
+\frac{2\Gamma(\hat{a})
}{\Gamma(\hat{a}-2)}\Big[\ln(2iu)+\psi(\hat{a})-\psi(1)-\psi(3)
\Big]u^2\right.
\nonumber\\
&&  \left.\qquad\qquad+\frac{2e^{-iu}u^2\Gamma(\hat{a})
}{\Gamma(\hat{a}-2)}\sum_{n=1}^\infty\frac{(\hat{a})_n(2iu)^n}{(3)_nn!}
\Big[\ln(2iu)+
\psi(\hat{a}+n)-\psi(1+n)-\psi(3+n)\Big]\right\}.
\end{eqnarray}

\subsection{The exact solution for the Klein-Gordon equation}

On the other hand, the equation of the motion for a massless scalar
field (\ref{k-g}) can be expressed as
\begin{eqnarray}\label{perturbation2}
\frac{1}{r^{3}}\frac{d}{dr} \left[ r^{3} f(r) \frac{d}{dr} \phi
\right] +\Big[
\frac{\omega^2}{(\mathrm{R}r)^{2}f(r)}-\frac{l(l+1)}{r^2}\Big]\phi =
0 \ .
\end{eqnarray}
 In order to solve the wave equation exactly, we take
\begin{eqnarray}
&&z=1-\frac{2u}{\tilde{\omega}}=1-\frac{r_H^2}{r^2},\ \nonumber  \\
&&\phi(z)=z^{\frac{i\tilde{\omega}}{2}}F(z).
\end{eqnarray}
Then Eq. (\ref{perturbation2}) becomes
\begin{eqnarray}
z(1-z)\frac{d^2F(z)}{dz^2}+(1+i\tilde{\omega}-i\tilde{\omega}z)
\frac{dF(z)}{dz}+\big[\frac{1}{2}i\tilde{\omega}-\frac{\tilde{L}^2}{4}\big]F(z)=0.
\end{eqnarray}
It is a hypergeometric differential equation. An exact solution of
the equation is
\begin{eqnarray}\label{333}
\phi(z)=z^{i\tilde{\omega}/2}\big[C_1F(a,b;c;z)+
C_2z^{1-c}F(a-c+1,b-c+1;2-c;z)\big],
\end{eqnarray}
with
\begin{eqnarray}c&=&1+i\tilde{\omega},\nonumber  \\
a&=&(i\tilde{\omega}-1+\sqrt{1-\tilde{L}^2-
\tilde{\omega}^2})/2,\ \nonumber  \\
b&=&(i\tilde{\omega}-1-\sqrt{1-\tilde{L}^2-
\tilde{\omega}^2})/2.\nonumber
\end{eqnarray}

\subsection{The greybody factor}

In the near horizon region $r\rightarrow r_H$,  the asymptotic
results of Eq. (\ref{333}) can be written as
\begin{eqnarray}\label{wavehor}
\phi(r\rightarrow r_H )=C_1e^{i\omega x}+C_2e^{-i\omega x}.
\end{eqnarray}
Comparing with Eq. (\ref{solhor}), we have $  C_2=0$.  Therefore,
the solution of the Eq. (\ref{perturbation2}) is
\begin{eqnarray}\label{asy2}
\phi(z)=z^{i\tilde{\omega}/2}C_1F(a,b;c;z).
\end{eqnarray}

In the far-region, $z\rightarrow 1$, the hypergeometric function can
be expressed as \cite{abram}
\begin{eqnarray}
&&F(a,b;c;z)=\frac{\Gamma (2)\Gamma(c)}{\Gamma
(a+2)\Gamma(b+2)}\sum^{1}_{n=0}\frac{(a)_n(b)_n}{n!(-1)_n}(1-z)^n
-\frac{\Gamma(c)(1-z)^2}{\Gamma
(a)\Gamma(b)}\sum^{\infty}_{n=0}\frac{(a+2)_n(b+2)_n}{n!(n+2)!}(1-z)^n\nonumber\\
&&\qquad\qquad\qquad\times\{\psi(a+2+n)+\psi(b+2+n)-\psi(1+2+n)-\psi(1+n)+\ln(1-z)\}.
\end{eqnarray}
 Thus, as $z\rightarrow1$,
by using the expansion in the neighbor of the point $z=1$
 \begin{eqnarray} z^{\frac{i\tilde{\omega}}{2}}=1-\frac{i
 \tilde{\omega}}{2}(1-z)-
 (\frac{i\tilde{\omega}}{4}+\frac{ \tilde{\omega}^2}{8})(1-z)^2
 +\cdot\cdot\cdot\ ,
\end{eqnarray}
 Eq. (\ref{asy2}) can be
rewritten as
\begin{eqnarray}\label{asymptotic2}
\phi(z)&=&C_1\frac{4\Gamma(c)}{\tilde{\omega}^2\Gamma
(a)\Gamma(b)}\left\{\frac{4\tilde{\omega}^2}{4\tilde{\omega}^2+\tilde{L}^4}
\Big[1-\frac{\tilde{L}^2}{4}(1-z)-\frac{\tilde{\omega}^2}{8}(1-z)^2\Big(-1
+\frac{2i}{\tilde{\omega}}-\frac{i\tilde{L}^2}{\tilde{\omega}}\Big)\Big]
\right.
\nonumber\\
&&  \left. -\frac{\tilde{\omega}^2}{8}(1-z)^2
\Big[\psi(a+2)+\psi(b+2)-\psi(3)-\psi(1)+\ln(1-z)\Big]-\frac{\tilde{\omega}^2}{4}
\sum^{\infty}_{n=1}\frac{(a+2)_n(b+2)_n}{n!(n+2)!}\right.
\nonumber\\
&&  \left.
\cdot z^{\frac{i\tilde{\omega}}{2}}(1-z)^{n+2}
\Big[\psi(a+2+n)+\psi(b+2+n)-\psi(1+2+n)-\psi(1+n)+\ln(1-z)
\Big]\right\}.
\end{eqnarray}

By matching the far-region boundary condition (\ref{sol_asy2}) onto
Eq. (\ref{asymptotic2}), we obtain
\begin{eqnarray}
 \hat{C}_1&=& -C_1A(\omega)
\Big[\frac{4\tilde{\omega}^2}{4\tilde{\omega}^2+\tilde{L}^4}
\Big(C(\omega)+\frac{i}{\tilde{\omega}}\Big)
+\frac{B(\omega)}{2}\Big],\ \nonumber\\
 \hat{C}_2&=& -C_1A(\omega)\frac{16\Gamma(\hat{a})\tilde{\omega}^2}{4\tilde{\omega}^2+\tilde{L}^4},
  \label{factor2}
\end{eqnarray}
with
\begin{eqnarray}
A(\omega)&=&\frac{4\Gamma(c)}{\tilde{\omega}^2\Gamma
(a)\Gamma(b)},\ \nonumber\\
B(\omega)&=&\psi(a+2)+\psi(b+2)-\psi(3)-\psi(1)
,
 \nonumber\\
 C(\omega)&=&
 \frac{2\Gamma(\hat{a})
}{\Gamma(\hat{a}-2)}\Big[\frac{i\pi}{2}+\ln\tilde{\omega}+\psi(\hat{a})-\psi(1)-\psi(3)
\bigg].
  \label{factor3}
\end{eqnarray}

We now separate the far-region solution into the parts of outgoing
and ingoing waves. Eq. (\ref{asymptotic2}) becomes
\begin{eqnarray}
 \phi(u)&=& f_0\left\{\Big[\frac{2\hat{C}_1}{f_0}+2i\hat{C}_2\textrm{Im}\Big(\frac{f_1}{f_0}\Big) +2\hat{C}_2\textrm{Re}\Big(\frac{f_1}{f_0}\Big)\Big]\frac{ u^2}{2}+\frac{\hat{C}_1}{f_0}e^{-iu}u^2\sum_{n=1}^\infty\frac{(\hat{a})_n}{(3)_n}(2iu)^n\right.
\nonumber\\
&&  \left.+\hat{C}_2
 \Big[1-\frac{i\tilde{L}^2}{2\tilde{\omega}}u +\frac{2\Gamma(\hat{a})
}{\Gamma(\hat{a}-2)}u^2\ln\frac{2u}{\tilde{\omega}}\Big]
\right.\nonumber\\&&\left.
+\hat{C}_2e^{-iu}u^2\frac{2\Gamma(\hat{a})
}{\Gamma(\hat{a}-2)}\sum_{n=1}^\infty\frac{(\hat{a})_n(2iu)^n}{(3)_nn!}
\Big[\ln(2iu)+
\psi(\hat{a}+n)-\psi(1+n)-\psi(3+n)\Big]\right\}
 ,
\end{eqnarray}
with
\begin{eqnarray}
f_0&=&-\frac{1}{4\Gamma(\hat{a})},\nonumber \\
f_1&=&-\frac{1}{4\Gamma(\hat{a})}
\Big[C(\omega)+\frac{1}{2}+\frac{i\tilde{L}^2}{2\tilde{\omega}}\Big].\nonumber
\end{eqnarray}
Defining quantities $D_1$ and $D_2$ as
\begin{eqnarray}
&& D_1+D_2=\frac{2\hat{C}_1}{f_0}+2i\hat{C}_2\textrm{Im}
\Big(\frac{f_1}{f_0}\Big),
\\ && i(D_1-D_2)=2\hat{C}_2\textrm{Im}\Big(\frac{f_1}{f_0}\Big),
\end{eqnarray}
and using the tortoise coordinate in the far-region
$\frac{d}{dx}=-\omega\frac{d}{du}$, the asymptotic  flux
($u\rightarrow0$) is given by
\begin{eqnarray}
J_{asy}=4\pi r^2\cdot \frac{u\omega}{2\textrm{Im}
\big(\frac{f_1}{f_0}\big)} f_0f_0^*\Big(|D_2|^2-|D_1|^2\Big).
\end{eqnarray}
Its incoming and outgoing fluxes are
\begin{eqnarray} && J_{in}
=\frac{\pi\tilde{\omega}^2}{\textrm{Im}\big(\frac{f_1}{f_0}\big)}
\mathrm{R}^3r_H^4 |D_2|^2, \\ &&J_{out} =
\frac{\pi\tilde{\omega}^2}{\textrm{Im}\big(\frac{f_1}{f_0}
\big)}\mathrm{R}^3r_H^4|D_1|^2 \ . \label{flux-asy2}
\end{eqnarray}
Lastly, the greybody factor is
\begin{eqnarray}\label{jin} \gamma(\tilde{\omega})=1-\frac{J_{out}}{J_{in}}
=1-
\frac{\big|\frac{\hat{C_1}}{f_0}\big|^{2}}{\big|\frac{\hat{C_1}}{f_0}
+2i\hat{C_2}\textrm{Im}\big(\frac{f_1}{f_0}\big)\big|^{2}},
\end{eqnarray}
which is valid for any frequency. It is well known that for usual
black holes in Einstein's general relativity, one can only calculate
the greybody factors analytically in the low frequency region by
some approximations as the ones first set out in \cite{page,unruh},
or calculate them by numerical method \cite{jorge,kanti1,kanti3}.
Some notable exceptions appear in string theory when considering the
propagation of a massless scalar field in the background of an
extremal dyonic string in six dimensions \cite{cvetic}, and in the
background supergravity for the D3-brane \cite{gubser}, where one
can obtain analytical expressions for the greybody factor in terms
of Mathieu functions. For \textrm{3D} dilaton black holes, the
greybody factors can be analytically obtained only when frequency
$\omega\geq 2$ \cite{fernando}. Therefore, in four dimensional
spacetime our analytical result obtained here is very special one.

In order to have an intuitional picture, we now plot some figures
for $ \gamma(\tilde{\omega})$ with different angular numbers in
following.

{\it i)  The greybody factor for $s$-wave, i.e. the angular number
$l=0$:}

\begin{figure}[ht]
\begin{center}
\includegraphics[width=6cm]{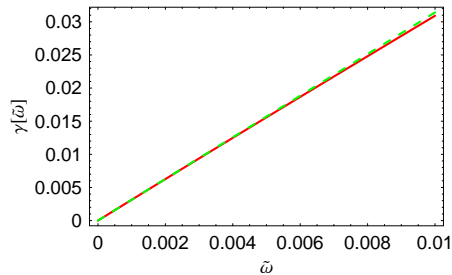}
 \ \;\;\;
\caption{(Color online) The greybody factor $\gamma(\tilde{\omega})$
of the $\lambda=1/2$ Ho\v{r}ava-Lifshitz black hole when
$\tilde{\omega}\in[0,0.01]$ for $s$-wave. The solid (red) line
present the greybody factor $\gamma(\tilde{\omega})$ described by
Eq. (\ref{jin}), and the dashed (green) line shows that
$\gamma(\tilde{\omega})\sim \pi \tilde{\omega}$ .\label{figdi}}
\end{center}
\end{figure}

\begin{figure}[ht]
\begin{center}
\includegraphics[width=6cm]{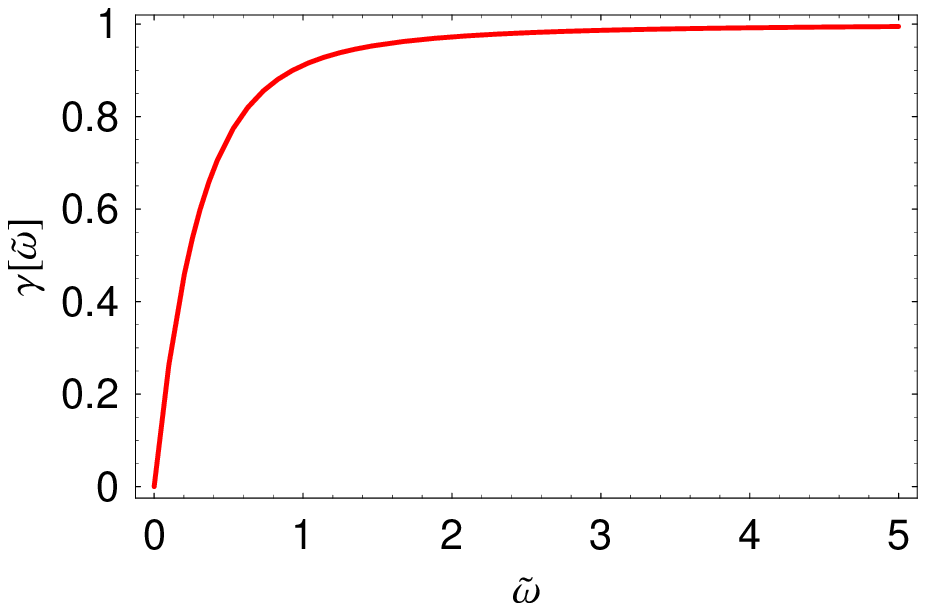}
\includegraphics[width=6cm]{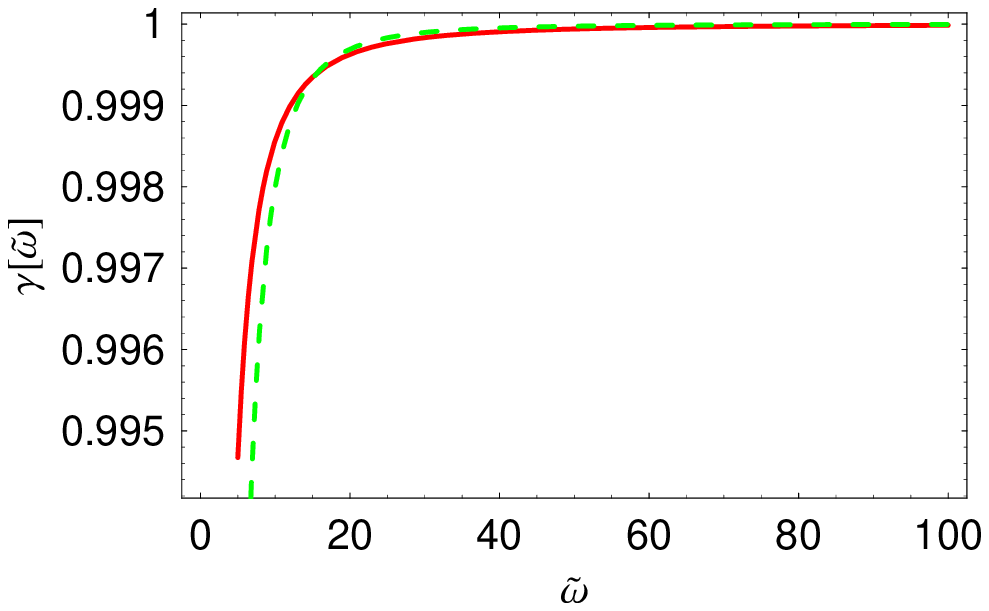}
\includegraphics[width=6cm]{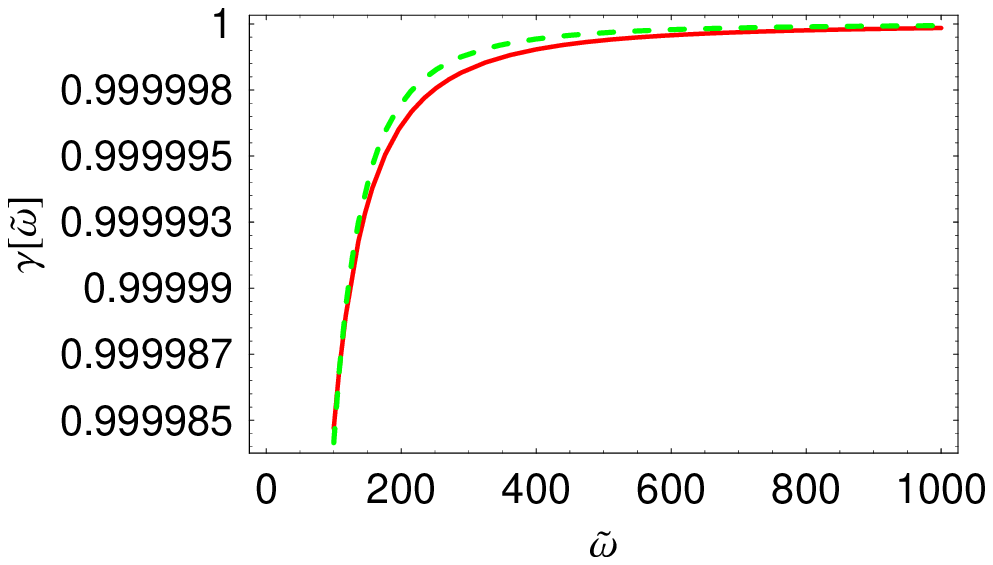}
\includegraphics[width=6cm]{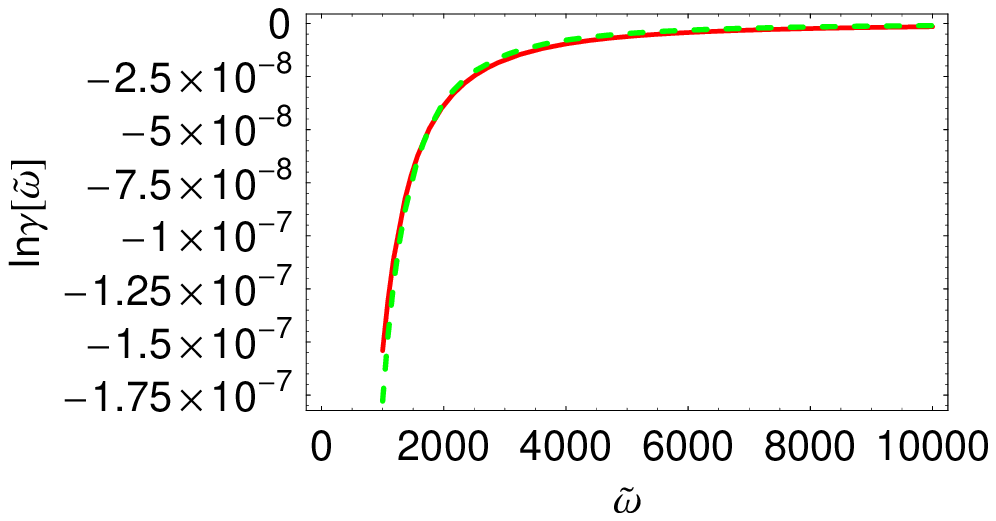}
 \ \;\;\;
\caption{(Color online) The greybody factor $\gamma(\tilde{\omega})$
of the $\lambda=1/2$ Ho\v{r}ava-Lifshitz black hole for $s$-wave.
The solid (red) line present the greybody factor
$\gamma(\tilde{\omega})$ described by Eq. (\ref{jin}). The dashed
(green) line in the second graph shows that
$\gamma(\tilde{\omega})\sim 1-\tilde{\omega}^{-2.7}$ when
$\tilde{\omega}\in[5,100]$; the dashed line in the third graph shows
that $\gamma(\tilde{\omega}) \sim 1-\tilde{\omega}^{-2.4}$ when
$\tilde{\omega}\in[100,1000]$; the last figure shows the logarithm
of the greybody factor $\ln\gamma(\tilde{\omega})$, and the dashed
line shows that
 $\gamma(\tilde{\omega})\sim 1-\tilde{\omega}^{-2.25}$ when $\tilde{\omega}\in[1000,10000]$.}\label{figd}
\end{center}
\end{figure}

We now show the greybody for $s$-wave in Figs. (\ref{figdi}) and
(\ref{figd}), which tell us that the curves of the greybody factor
in the $\lambda=1/2$ Ho\v{r}ava-Lifshitz spacetime are continuous in
all frequencies and it approaches to unity when
$\tilde{\omega}\geq5$ which is similar to the numerical results for
the black 3-brane in
 an asymptotically AdS$_5\times$S$^5$ space \cite{jorge}.

Fig. (\ref{figdi}) shows that the greybody
factor $\gamma(\tilde{\omega})\sim \pi \tilde{\omega}$ in the low frequency region when the
frequency of the scalar field $\tilde{\omega}\in(0,0.01]$. For asymptotically AdS spacetime,
the result is $\gamma(\tilde{\omega})\sim  \tilde{\omega}^2$ \cite{harmark} by using
the low frequency approximation approach ($\tilde{\omega}\ll 1$).
For the black 3-brane in
 an asymptotically AdS$_5\times$S$^5$ space, the result is $\gamma(\tilde{\omega})\sim  \tilde{\omega}^3$ \cite{jorge} via the low frequency approximation method ($\tilde{\omega}\ll 1/\pi$).

 Fig.
(\ref{figd}) gives that $\gamma(\tilde{\omega})\sim
1-\tilde{\omega}^{-2.7}$ when $\tilde{\omega}\in[5,100]$;
$\gamma(\tilde{\omega})\sim 1-\tilde{\omega}^{-2.4}$ when
$\tilde{\omega}\in[100,1000]$; and $\gamma(\tilde{\omega})\sim
1-\tilde{\omega}^{-2.25}$ when $\tilde{\omega}\in[1000,10000]$.
For the black 3-brane in
 an asymptotically AdS$_5\times$S$^5$ space, the result is $\gamma(\tilde{\omega})\sim  1-2.3\tilde{\omega}^{-8}$ when $\tilde{\omega}\in[8,15]$ by using numerical scheme \cite{jorge}.

{\it ii) The greybody factor for the angular number $l\neq0$:}

\begin{figure}[ht]
\begin{center}
\includegraphics[width=6cm]{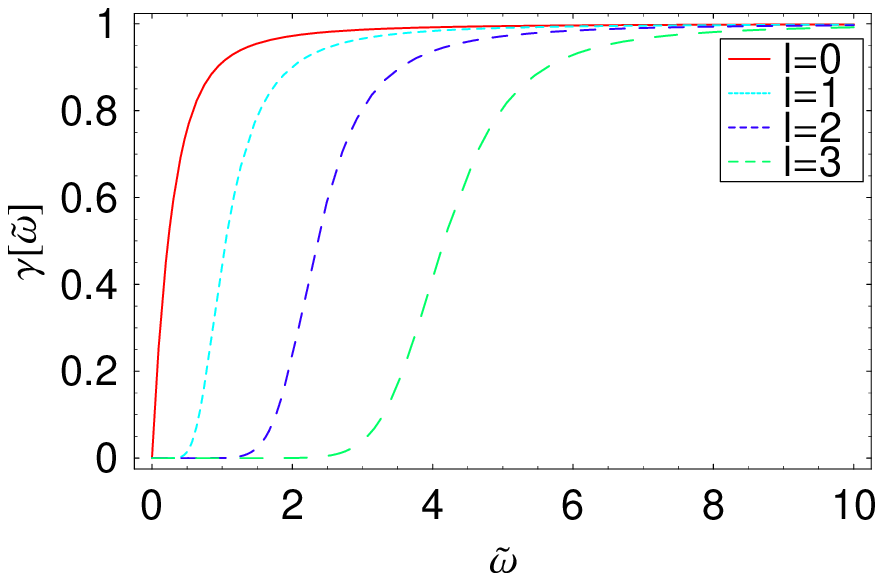}
\includegraphics[width=6cm]{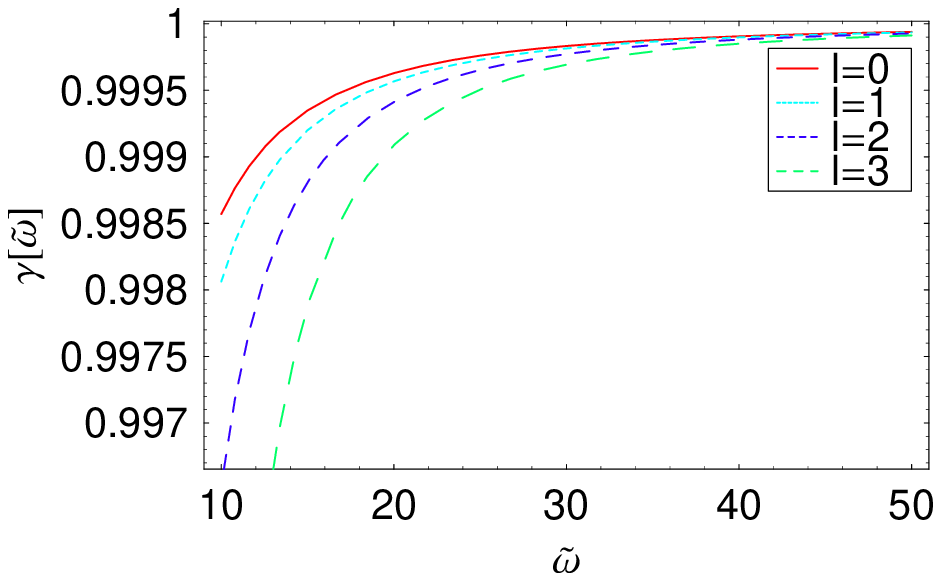}
 \ \;\;\;
\caption{(Color online) The greybody factor $\gamma(\tilde{\omega})$
of the $\lambda=1/2$ Ho\v{r}ava-Lifshitz black hole with different
angular number $l$. The lines from left to right in two graphs are
corresponding to $l=0,~1,~2,~3$.}\label{figg2}
\end{center}
\end{figure}

The curves in Fig. (\ref{figg2}) show the greybody factor with
different angular number $l$ in the $\lambda=1/2$
Ho\v{r}ava-Lifshitz black-hole spacetime. It is easy to see that the
greybody factor decreases as the  angular number $l$ increases, and
approaches to unity at large frequencies. These features are similar
to usual black holes \cite{kanti1,kanti3,kanti4,sbchen}.

\section{Dynamic evolution of the scalar field perturbation
using analytical method}

\begin{figure}[ht]
\begin{center}
\includegraphics[width=6cm]{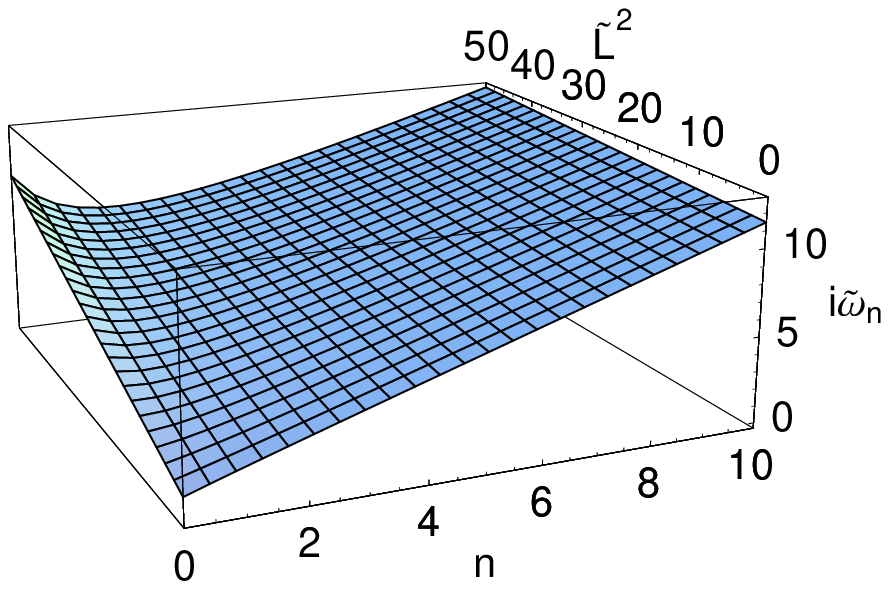}
\includegraphics[width=6cm]{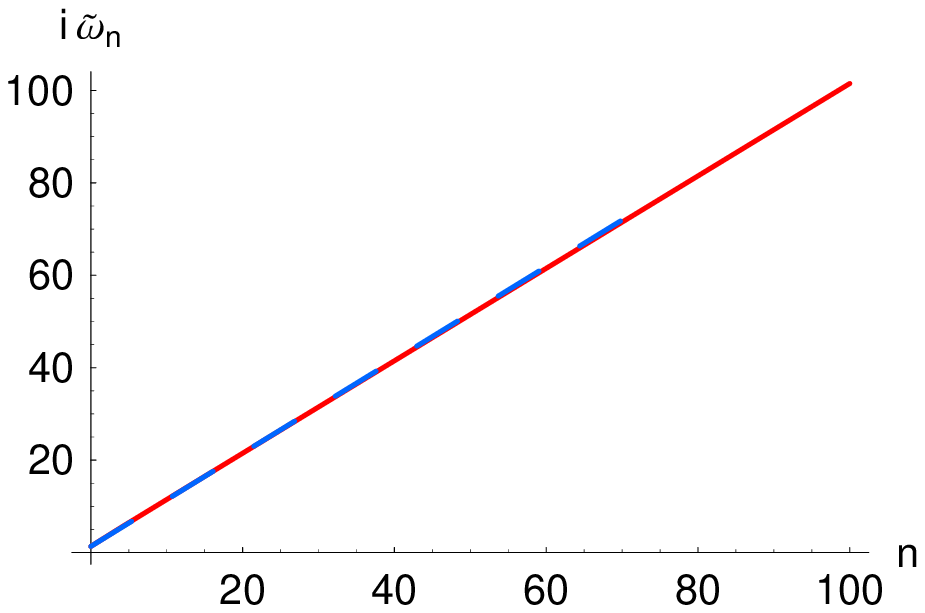}
 \ \;\;\;
\caption{(Color online) The left graph is the asymptotic behavior of
high overtones $\tilde{\omega}_n$ of purely damped frequency of the
scalar field perturbation versus overtone number $n$ and different
angular number $l$. The right graph shows that the analytical result
(solid (red) line) and the numerical result (dashed (blue) line)
when $l=0$. }\label{figd2}
\end{center}
\end{figure}

In Ref. \cite{ding}, we found that the dynamic evolution of the
scalar field perturbation in the $\lambda=1/2$ Ho\v{r}ava-Lifshitz
black-hole spacetime is purely damped modes for $l=0$ by using
numerical method. In this section, we show that we can obtain
dynamic evolution modes from Eq. (\ref{333}) by using analytical
method for any $l$. The quasinormal modes of a classical
perturbation of black-hole spacetime are defined as the solutions of
the wave equations with purely ingoing waves at the horizon, and
the amplitude of the ingoing wave has to be zero at the asymptotic
region $r\rightarrow\infty$. This leads to $|D_2|^2=0$, i.e.,
$A(\omega)=0$. Hence, Eq. (\ref{factor2}) shows us that the quantity
$a+2$ or $ b+2$ should be a negative integer
\begin{eqnarray} a+2= \frac{1}{2}(i\tilde{\omega}-1+\sqrt{1-\tilde{L}^2-\tilde{\omega}^2})
+2=-n, ~~~~ (n=0,\ 1,\ 2,\ 3,\  \cdot\cdot\cdot).
\end{eqnarray} Solving it we obtain
\begin{eqnarray} \tilde{\omega}_n=-i\left[n+\frac{3}{2}-
\frac{1-\tilde{L}^2}{2(2n+3)}\right], \label{qnm}
\end{eqnarray}
which is showed in Fig. (\ref{figd2}). If we take
$l=0$ and use the equation $\tilde{\omega}=\omega/(2\pi
T_H^{\lambda=1/2})$, this analytical result coincides with  the numerical  results
$\tilde{\omega}_n=-i(1.009n+1.333)$ in Ref. \cite{ding} (see Fig. \ref{figd2}). These modes are purely
damped modes which may be caused by its over-damped potential, and
these purely damped modes show that this spacetime is very stable.

\section{Summary and discussion}

We have studied the propagation and dynamic evolution of the
massless scalar field  in the background of the $\lambda=1/2$
Ho\v{r}ava-Lifshitz black hole with the  condition of the detailed
balance. It is interesting to note that we obtained an exact
solution for the Klein-Gordon equation in the whole spacetime. Then,
imposing the boundary conditions at the event horizon and infinity,
we have found an analytical expression (\ref{jin}) for the greybody
factor of the scalar field propagating in the Ho\v{r}ava-Lifshitz
black-hole spacetime. The greybody factor approaches to unity when
$\tilde{\omega}\geq5$ and is valid both for any frequency and
angular  number. To compare with other results \cite{jorge,harmark},
we have also shown behaviors of the greybody factor for the
low-frequency and high-frequency for $l=0$: the greybody factor
$\gamma(\tilde{\omega})\sim \pi \tilde{\omega}$ when the frequency
of scalar field $\tilde{\omega}\in(0,0.01]$;
$\gamma(\tilde{\omega})\sim 1-\tilde{\omega}^{-2.7}$ when
$\tilde{\omega}\in[5,100]$; $\gamma(\tilde{\omega})\sim
1-\tilde{\omega}^{-2.4}$ when $\tilde{\omega}\in[100,1000]$; and
$\gamma(\tilde{\omega})\sim 1-\tilde{\omega}^{-2.25}$ when
$\tilde{\omega}\in[1000,10000]$.

On the other hand, we have calculated the quasinormal modes of the
black hole by using analytical method. The exact result for the
modes is given by equation (\ref{qnm}). If we take $l=0$, then the modes
coincide with the numerical results
$\tilde{\omega}_n=-i(1.009n+1.333)$ in Ref. \cite{ding}.

\begin{acknowledgments}

This work was supported by the National Natural Science Foundation
of China under Grant No 10875040; the key project of the National
Natural Science Foundation of China under Grant No 10935013; the
National Basic Research of China under Grant No. 2010CB833004, the
Hunan Provincial Natural Science Foundation of China under Grant No.
08JJ3010,  PCSIRT under Grant No. IRT0964, and the Construct Program
of the National Key Discipline. S. Chen's work was partially
supported by the National Natural Science Foundation of China under
Grant No.10875041 and the construct program of key disciplines in
Hunan Province.
\end{acknowledgments}

\end{document}